\documentclass[prb, twocolumn]{revtex4}

\begin{document}

\title{
A lattice bosonic model as a quantum theory of gravity
}
\author{Zheng-Cheng Gu$^\dagger,^{\dagger\dagger}$ and  Xiao-Gang Wen$^\dagger$}
\affiliation{Department of Physics, Massachusetts Institute of
Technology, Cambridge, Massachusetts 02139$^\dagger$\\Center for
Advanced Study, Tsinghua University, Beijing, China,
100084$^{\dagger\dagger}$}
\date{{\small \today}}
\begin{abstract}
A local quantum bosonic model on a lattice is constructed whose low energy
excitations are gravitons described by linearized Einstein action.  Thus the
bosonic model is a quantum theory of gravity, at least at the linear level.
We find that the compactification and the discretization of metric tenor are
crucial in obtaining a quantum theory of gravity.
\end{abstract}

\maketitle

\emph{\bf Seven wonders of our universe}:
Our world has many mysteries and wonders.
At the most fundamental level,
there are may be seven deep mysteries and/or wonders in our universe:
(1) Identical particles.
(2) Gauge interactions.\cite{Wey52,P4103,YM5491}
(3) Fermi statistics.\cite{F2602,D2661}
(4) Tiny masses of fermions ($\sim 10^{-20}$ of the Planck
mass).\cite{GW7343,P7346,Wqoem}
(5) Chiral fermions.\cite{LY5654,Wo5713}
(6) Lorentz invariance.\cite{E0591}
(7) Gravity.\cite{E1669}
Here we would like to question where those wonderful and mysterious properties
come from. We wish to have a single unified understanding of all of the above
mysteries.  Or more precisely, we wish that we can start from a single
model to obtain all of the above wonderful properties.

Recent progresses
showed that, starting from a single origin -- a local bosonic model (which is
also called spin model), the first four properties in the above list emerge at
low energies.\cite{LWuni,Wqoem,LWqed,Wbynd}  Thus we may say that the local
bosonic model provides a unified origin for identical particles, gauge
interactions, Fermi statistics and near masslessness of the fermions.  Gauge
interaction and Fermi statistics are unified under this point of view of
emergence.
However, three more mysteries remain to be understood.
In this paper,
we will
show that it is possible to construct a local bosonic model from which
gravitons emerge.  On one hand, such a model can be viewed as a theory of
quantum gravity.  It solves a long standing problem of putting quantum
mechanics and gravity together.  On the other hand, the model provides a design
of a condensed matter system which has an emergent quantum gravity at low
energies.  This allows us to study some quantum effects of gravity (a simulated
one) in a laboratory.

The belief that all the wonderful phenomena of our universe (such as
gauge interaction, Fermi statistics, and gravitons) emerge from a
lowly local bosonic model is called locality principle.\cite{Wqoem}
Using local bosonic model as a underlying structure to understand
the deep mysteries of our universe represent a departure from the
traditional approach of gaining a better understanding by dividing
things in to smaller parts.  In the traditional approach, we assume
that the division cannot continue forever and view everything as
made of some simple indivisible building blocks -- the elementary
particles.
However, the traditional approach may represent a
wrong direction.  For example, phonons behave just like any other elementary
particles at low energies.  But if we look at phonons closely, we do not see
smaller parts that form a phonon.  We see the atoms that form the
entire lattice.  The  phonons are not formed by those atoms, the phonons are
simply collective motions of those atoms.  This makes us to wonder that
photons, electrons, gravitons, \etc, may also be emergent phenomena just like
phonons. They may not be the building blocks of everything. They may be
collective motions of a deeper underlying structure.

This paper uses a particular emergence approach:
we try to obtain everything from a local bosonic model.
The detail form of the bosonic model is not important.
The important
issue is how the bosons (or the spins) are organized in the ground state.
It is shown that if bosons organize into a string-net condensed state, then
photons, electrons and quarks can emerge naturally as collective motions of
the bosons.\cite{LWuni,Wqoem,LWqed,Wbynd} In this paper, we will find an
organization of bosons such that the collective motions of bosons lead to
gravitons.

Other emergence approaches were developed in superstring
theory\cite{GreSW88} in last 10 years, as demonstrated by the
duality relations among various superstring models and matrix
models.\cite{Pol98} The anti-de Sitter-space/conformal-field-theory
duality even shows how space-time and gravity emerge from a gauge
theory.\cite{AGM0083}

\emph{\bf The rule of game}: There are many different approaches to quantum
gravity\cite{T0294,C0185,P9487} based on different principles. Some
approaches, such as loop quantum gravity,\cite{S0448} stress the gauge
structure from the diffeomorphism of the space-time.  Other approaches, such
as superstring theory,\cite{GreSW88,Pol98} stress the renormalizability of the
theory.  In this paper, we follow a different rule of game by stressing
finiteness and locality.

Our rule of game is encoded in the following working
definition of quantum gravity.  Quantum gravity is\\
(a) a quantum theory.\\
(b) Its Hilbert space has a finite dimension.\\
(c) Its Hamiltonian is a sum of local operators.\\
(d)
The gapless helicity $\pm 2$ excitations are the only low energy excitations.\\
(e) The  helicity $\pm 2$ excitations have a linear dispersion.\\
(f) The gravitons \footnote{In this paper, we define gravitons as linearly
dispersing gapless helicity $\pm 2$ excitations.  }
can interact in the way consistent with experimental observations.

The condition (b) implies that the quantum gravity considered here has a
finite cut-off. So the renormalizability is not an issue here.  The condition
(c) is a locality condition.  It implies two additional things: (1) the total
Hilbert space is a direct product of local Hilbert spaces $\mathcal{H}_{tot} =
\otimes_n \mathcal{H}_n$,  (2) the local operators are defined as operators
that act within each local Hilbert space $\mathcal{H}_n$ or finite products of
local operators.  The conditions (a -- c) actually define a local bosonic
model.  Certainly, any quantum spin models satisfy (a -- c).  It is the
conditions (d--f) that make the theory to look like gravity.

The condition (d) is very important.
It is very easy to construct a quantum model that contains helicity
$\pm 2$ gapless excitations, such as the theory described by the following
Lagrangian $ \cL= \frac12 \prt_0 h^{ij}\prt_0 h^{ij} -\frac12 \prt_k
h^{ij}\prt_k h^{ij} $.
Such a theory is not a theory of gravity since it also contain helicity $0$,
$\pm 1$ gapless excitations.


Superstring theory\cite{GreSW88,Pol98} satisfies the conditions (a), (e) and
(f), but in general not (d) due to the presence of dilatons (massless scaler
particles).  The superstring theory (or more precisely, the superstring field
theory) also does not satisfy the condition (b) since the cut-off is not
explicitly implemented.  The spin network\cite{RS9543} or the quantum
computing\cite{S0535} approach to quantum gravity satisfies the condition
(a,b) or (a--c). But the properties (d--f) are remain to be shown.  The
induced gravity from superfluid $^3$He discussed in \Ref{V9867} does not
satisfy the condition (d) due to the presence of gapless density mode.  In
\Ref{ZH0123}, it is proposed that gravitons may emerge as edge excitations of
a quantum Hall state in 4 spatial dimensions. Again the condition (d) is not
satisfied due to the presence of infinite massless helicity $\pm 1$
,$\pm 2$, $\pm 3$, $\cdots$ modes.

In  \Ref{X0643}, a very interesting bosonic model is constructed which
contains gapless helicity $0$ and $\pm 2$ excitations with quadratic
dispersions.  The model satisfies (a--c), but not (d,e).  In this paper, we
will fix the two problems and construct a local bosonic model that satisfies the
conditions (a -- e) and possibly (f).  To quadratic order, the low energy
effective theory of our model is the linearized Einstein gravity. The key step
in our approach is to \emph{discretize and compactify} the metric tensor.

\emph{\bf Review of emergence of $U(1)$ gauge theory}: Our model for emergent
quantum gravity is closely related to the rotor model that produces emergent
$U(1)$ gauge theory.\cite{MS0204,Wen04} So we will first discuss the emergence
of $U(1)$ gauge theory to explain the key steps in our argument in a simpler
setting.

To describe the rotor model, We introduce an angular variable $a_{\v i\v
j}\sim a_{\v i\v j} +2\pi$ and the corresponding angular momentum $\cE_{\v i\v
j}$ for each link of a cubic lattice.  Here $\v i$ labels the sites of the
cubic lattice and $a_{\v i\v j}$ and $\cE_{\v i\v j}$ satisfy $a_{\v i\v
j}=-a_{\v j\v i}$ and $\cE_{\v i\v j}=-\cE_{\v j\v i}$. The phase space
Lagrangian for physical degrees of freedom $a_{\v i\v j}$ and $\cE_{\v i\v j}$
is given by\cite{Wen04}
\begin{align}
\label{lattLC} &L = \sum_{\<\v i\v j\>} \cE_{\v i\v j} \dot a_{\v
i\v j} -\frac J2 \sum_{\<\v i\v j\>} \cE_{\v i\v j}^2 +g\sum_{\<\v
i\v j\v k\v l\>} \cos\cB_{\v i\v j\v k\v l} -\frac U2 \sum_{\v i}
Q_{\v
i}^2\nonumber\\
&\cB_{\v i\v j\v k\v l} = a_{\v i\v j}+ a_{\v j\v k}+ a_{\v k\v l}+
a_{\v l\v i}, \quad Q_{\v i} = \sum_{\v j\text{ next to }\v i}
\cE_{\v i\v j}
\end{align}
where $\sum_{\v i}$ sums over all sites, $\sum_{\<\v i\v j\>}$
over all links, and $\sum_{\<\v i\v j\v k\v l\>}$ over all
square faces of the cubic lattice. We note that after quantization,
$\cE_{\v i\v j}$ are quantized as integers.

To obtain the low energy dynamics of the above rotor model,
let us assume that the fluctuation of $\cE_{\v i\v j}$ are large and treat
$\cE_{\v i\v j}$ as a continuous quantity.  We also assume that the
fluctuations of $a_{\v i\v j}$ are small and expand \eq{lattLC} to the
quadratic order of $a_{\v i\v j}$.  Then we take the continuum limit by
introducing the continuous fields $(\cE^i,a^i)$ and identifying $\cE_{\v i\v j}
=\int_{\v i}^{\v j}\dd  x^i \cE^i $ and $ a_{\v i\v j} =\int_{\v i}^{\v j}\dd
x^i a^i $. Here we have assumed that the lattice constant $a=1$.  The resulting
continuum effective theory is given by
\begin{equation}
\label{M4a}
 \mathcal{L}=
 - \mathcal{E}^i \partial_0a_i
-\frac{1}{2}J (\mathcal{E}^i)^2- \frac 12 g (\mathcal{B}^i)^2
-\frac12 U(\partial_i \mathcal{E}^i)^2,
\end{equation}
where $\cB^i=\eps^{ijk}\prt_j a_k$.
We find that the rotor model has three low lying modes. Two of them are
helicity $\pm 1$ modes with a linear dispersion $\om_{\v k} \sim \sqrt{gJ}|\v
k|$ and the third mode is the helicity $0$ mode with zero frequency $\om_{\v
k}=0$.  We know that a $U(1)$ gauge theory only have two helicity $\pm 1$
modes at low energies. Thus the key to understand the emergence $U(1)$ gauge
theory is to understand how the helicity $0$ mode obtain an energy gap.

To understand why helicity $0$ mode is gapped, let us consider the
quantum fluctuations of $\cE^i$ and $a_i$. We note that the
longitudinal mode and the transverse modes separate.  Introduce
$\cE^i=\cE_{||}+\cE_{\perp}$ and $a_i=a_{||}+a_{\perp}$, we find
that the dynamics of the transverse mode is described by
$\cL_{\perp}= a_{\perp}\prt_0 \cE_{\perp} -\frac J2 \cE_{\perp}^2 -
\frac g2 \prt_i a_{\perp} \prt_i a_{\perp}$. At the lattice scale
$\del x\sim 1$, the quantum fluctuations of $\cE_{\perp}$ and
$a_{\perp}$ are given by $ \del \cE_{\perp}\sim \sqrt{\frac gJ}$ and
$\del a_{\perp}\sim \sqrt{\frac Jg}$.  We see that the assumptions
that we used to derive the continuum limit are valid when $J\ll g$.
In this limit we can trust the
result from the continuum effective theory and conclude that the
transverse modes (or the helicity $\pm 1$ modes) have a linear
gapless dispersion.

The longitudinal mode is described by $(f(\v x), \pi(\v x))$
with $a_i=\prt_i f$ and $\pi
=\prt_i\cE^i$. Its dynamics is determined by $\cL_{||}= \pi \prt_0 f
-\frac J2 \pi(-\prt^{-2})\pi -\frac 12 U \pi^2$. At the lattice
scale, the quantum fluctuations of $\pi$ and $f$ are given by $\del
\pi = 0$ and $\del f = \infty$.  We see that a positive $U$ and $J$
will make the fluctuations of $f$ much bigger than the
compactification size $2\pi$ and the fluctuations of $\pi$ much less
then the discreteness of $\cE^i$ which is $1$.  In this limit,
the result from the classical equation of motion cannot be trusted.

In fact the weak quantum fluctuations in the discrete
variable $\pi$ and the strong quantum fluctuations in the compact
variable $f$ indicate that the corresponding mode is gapped after
the quantization.  Since $\pi$ has weak fluctuations which is less
than the discreteness of $\pi$, the ground state is basically given
by $\pi=0$.
A low lying
excitation is then given by $\pi=0$ everywhere except in a unit cell
where $\pi = 1$.  Such an excitation have an energy of order $U$. The
gapping of helicity $0$ mode is confirmed by more
careful calculations.\cite{Wen04} From those calculations, we find
that the weak fluctuations of $\pi$ lead to a constraint $
\pi=\prt_i\cE^i=0$ and the strong fluctuations of $f$ lead to a
gauge transformation $ a_i\to a_i+\prt_i f$. The Lagrangian \eq{M4a}
equipped with the above constraint and the gauge transformation
becomes the Lagrangian of a $U(1)$ gauge theory.  We will use
this kind of argument to argue the emergence of gravitons and
the linearized Einstein action.

\emph{\bf The emergence of quantum gravity}:
First, let us describe a model that will have emergent gravitons at low
energies.  The model has six variables $\th_{xx}(\v i)$, $\th_{yy}(\v i)$,
$\th_{zz}(\v i)$, $L^{xx}(\v i)$, $L^{yy}(\v i)$, and $L^{zz}(\v i)$ on
each vertex of a cubic lattice.  The model also has two  variables on each
square face of the cubic lattice. For example, on the square centered at $\v
i+\frac{\v x}{2} +\frac{\v y}{2}$, the two variables are
$
\th_{xy}(\v i+\frac{\v x}{2} +\frac{\v y}{2})
=
\th_{yx}(\v i+\frac{\v x}{2} +\frac{\v y}{2})
$ and $
L^{xy}(\v i+\frac{\v x}{2} +\frac{\v y}{2})
=
L^{yx}(\v i+\frac{\v x}{2} +\frac{\v y}{2})
$.
The bosonic model is described by the following phase space Lagrangian
\begin{align}
\label{LagSp}
 L& =
\sum_{\v i,ab=xy,yz,zx} L^{ab}(\v i+\frac{\v a}{2} +\frac{\v b}{2}) \prt_0 \th_{ab}(\v i+\frac{\v a}{2} +\frac{\v b}{2})
\nonumber\\
&+\sum_{\v i,a=x,y,z} L^{aa}(\v i) \prt_0 \th_{aa}(\v i)
-H_U-H_J-H_g
\end{align}
where
\begin{align}
H_U =& n_G U_1\sum_{\v i} \sum_{a=x,y,z} \{1-\cos[2\pi Q(\v i,\v
i+\v a)/n_G]\} \nonumber\\+&n_G U_2\sum_{\v i} \{1-\cos[\eta(\v
i)]\} \nonumber\\
  H_J =&n_G  J\sum_{\v i,a=x,y,z} \{1-\cos[2\pi L^{aa}(\v i)/n_G]\}
  \nonumber\\+& 2n_G  J\sum_{\v i}\sum_{ab=xy,yz,zx} \{1-\cos[2\pi L^{ab}(\v i+\frac{\v a}{2}+\frac{\v b}{2})/n_G]\}
  \nonumber\\-& \frac12 n_G  J\sum_{\v i}\{1-\cos[2\pi\sum_{a=x,y,z} L^{aa}(\v i)/n_G]\}
\nonumber\\
 H_g = &\frac{n_G  g}{4}\sum_{\v i,a=x,y,z} \{1-\cos[\rho^a_a(\v i)]\}
 \nonumber\\+&\frac{n_G  g}{4}\sum_{\v i,ab=xy,yz,zx} \sin[\rho^a_b(\v i)]\sin[\rho^b_a(\v
 i)]\}
\end{align}
Here $\rho^i_j(\v i)$, $\eta(\v i)$ and $Q(\v i,\v i+\v x)$ are
defined as $ \rho^x_x(\v i)  =
  \th_{zx}(\v i+\v y +\frac{\v z}{2}+\frac{\v x}{2})
+ \th_{zx}(\v i+\frac{\v z}{2}+\frac{\v x}{2})
- \th_{xy}(\v i+\v z +\frac{\v x}{2}+\frac{\v y}{2})
- \th_{xy}(\v i +\frac{\v x}{2}+\frac{\v y}{2})$,
$\
\rho^x_y(\v i)  = - \th_{yz}(\v i +\frac{\v z}{2}+\frac{\v y}{2}) -
\th_{yz}(\v i+\frac{\v z}{2}-\frac{\v y}{2})
+2\th_{yy}(\v i+\v z)
+2\th_{yy}(\v i)
$,
$\
\rho^x_z(\v i) =
-2\th_{zz}(\v i+\v y)
-2\th_{zz}(\v i)
+ \th_{yz}(\v i+\frac{\v y}{2}+\frac{\v z}{2})
+ \th_{yz}(\v i+\frac{\v y}{2}-\frac{\v z}{2})
$,
$\
 \eta(\v i) = \ \sum_{ a= x, y, z}\sum_{b=x,y,z}
[\th_{bb}(\v i+\v  a) +\th_{bb}(\v i-\v  a) +2\th_{bb}(\v i)]
-  \sum_{ a= x, y, z}
[\th_{ a a}(\v i+\v  a) +\th_{ a a}(\v i-\v  a) +2\th_{ a a}(\v i)]
-  \sum_{ab=xy,yz,zx}
[\th_{ab}(\v i+\frac{\v a}{2}+\frac{\v b}{2})
+\th_{ab}(\v i-\frac{\v a}{2}+\frac{\v b}{2})
+\th_{ab}(\v i+\frac{\v a}{2}-\frac{\v b}{2})
+\th_{ab}(\v i-\frac{\v a}{2}-\frac{\v b}{2})],
$
and
$\
 Q(\v i,\v i+\v x) =
 L^{xx}(\v i+\v x) + L^{xx}(\v i)
+L^{yx}(\v i+\frac{\v x}{2}+\frac{\v y}{2})
+L^{yx}(\v i+\frac{\v x}{2}-\frac{\v y}{2})
+L^{zx}(\v i+\frac{\v x}{2}+\frac{\v z}{2})
+L^{zx}(\v i+\frac{\v x}{2}-\frac{\v z}{2})$.
Other components are obtained by cycling $xyz$ to $yzx$ and $zxy$.

Note that both $\th_{ab}$ and its canonical conjugate $L^{ab}$ are
compactified: $ \th_{ab}\sim \th_{ab}+2\pi $, $ L^{ab}\sim L^{ab}+n_G$.  Hence
after quantization they are both discretized and $n_G$ is an integer. Due to
the compactification, only $W_L^{ab}=\e^{2\pi \imth L^{ab}/n_G}$,
$W_\th^{ab}=\e^{\imth \th_{ab}}$ and their products are physical operators.
For a fix $ab$ and $\v i$, $W_L^{ab}(\v i)$ and $W_\th^{ab}(\v i)$ satisfy the
algebra
\begin{equation}
\label{WW}
 W_L^{ab}(\v i) W_\th^{ab}(\v i)
= \e^{2\pi \imth/n_G} W_\th^{ab}(\v i) W_L^{ab}(\v i)
\end{equation}
Such an algebra has only one $n_G$ dimensional representation.  This $n_G$
dimensional representation becomes our local Hilbert space $\cH_{\v i,ab}$. The
total Hilbert space of our model \eq{LagSp} is given by $\cH=\otimes_{\v i,ab}
\cH_{\v i,ab}$ after quantization.  In other words, there are $n_G^3$ states on
each vertex and $n_G$ states on each square face of the cubic lattice.
Note that the Hamiltonian
\begin{equation}
H=H_U+H_J+H_g \label{HUJg}
\end{equation} is a function of the physical operators $W_L^{ab}$,
$W_\th^{ab}$ and their hermitian conjugates.
So the quantum model defined through the Hamiltonian \eq{HUJg} and the algebra
\eq{WW} is a bosonic model whose local Hilbert spaces have finite dimensions.

Next, we would like to understand low energy excitations of the quantum bosonic model \eq{HUJg} in large $n_G$ limit.
We first assume that the fluctuations of $\phi^{ij}\equiv 2\pi L^{ij}/n_G$ and
$\th_{ij}$ are much bigger then $1/n_G$ (the discreteness of $\phi^{ij}$ and
$\th_{ij}$) so that we can treat $\phi^{ij}$ and
$\th_{ij}$ as continuous variables.  We also assume that the fluctuations of
$\phi^{ij}$ and $\th_{ij}$ are much smaller then $1$ so that we can treat
$\phi^{ij}$ and $\th_{ij}$ as small variables. Under those assumptions, we can
use semiclassical approach to understand the low energy dynamics of the quantum
bosonic model \eq{HUJg}.

Expanding the Lagrangian \eq{LagSp} to quadratic order in  $\phi^{ij}$ and
$\th_{ij}$, we can find the dispersions of collective modes of the bosonic
model.
There are total of six collective modes. We find four of them have zero
frequency for all $\v k$, and two modes have a linear dispersion relation near
$\v k=(\pi,\pi,\pi)$. Near $\v k=(\pi,\pi,\pi)$, the dynamics of the six
modes are described by the following continuum field theory:
\begin{align}
\label{Lagthphi}
\cL & =n_G\Big\{
\phi^{ij} \dot \th_{ij}
-\frac{J}{2}\Big[(\phi^{ij})^2 -\frac{(\phi^{ii})^2}{2}\Big]
-\frac{g}{2} \th_{ij}R^{ij}
\nonumber\\
&\ \ \ \ \ \ \ \ \ \ -\frac{U_1}{2}(\prt_i\phi^{ij})^2
-\frac{U_2}{2} (R^{ii})^2 \Big\}
\end{align}
where $R^{ij}=\eps^{imk}\eps^{jln}\prt_m\prt_l \th_{nk}$ and we define the
continuum field $\theta^{ab}(\v x)$ as $-(-1)^{\v i}\frac12 \theta^{ab}(\v
i+\frac{\v a}{2}+\frac{\v b}{2})$ for $a\neq b$ and as $(-1)^{\v i}
\theta^{ab}(\v i)$ for $a=b$.
The helicity $\pm 2$ modes have a linear
dispersion relation $\om_{\v k} \sim \sqrt{gJ} |\v k|$. We find that
for large $n_G$, the quantum fluctuations of the helicity $\pm 2$
modes is of order $\del \phi^{ij},\del\th_{ij}\sim \sqrt{1/n_G}$
(assuming $ U_{1,2}$, $ J$ and $ g$ are of the same order). So the
fluctuations of $\phi^{ij}$ and $\th_{ij}$ satisfy $1/n_G\ll \del
\phi^{ij},\del \th_{ij}\ll 1$ and the semiclassical approximation is
valid for the helicity $\pm 2$ modes. In this case, the result
$\om_{\v k}\sim \sqrt{gJ}|\v k|$ can be trusted.

The helicity $\pm 1$ modes and one of the helicity $0$ mode are described by
$\th_{ij}= \prt_i \th_j+\prt_j \th_i$ and $\phi^i=\prt_j \phi^{ji}$.  Their
frequency $\om_{\v k}=0$.  For such modes,
the Hamiltonian only contains $\phi^i$. Thus the
quantum fluctuations satisfies $\del \phi^{i} \ll 1/n_G$ and $\del \th_{i}\gg
1$. So the semiclassical approximation is not valid and the result $\om_{\v
k}=0$ cannot be trusted.  Using the similar argument used in emergence of
$U(1)$ gauge bosons, we conclude that those modes are gapped.  The strong
fluctuations $\del \th_{ij}= \prt_i \th_j+\prt_j \th_i \gg 1$
and the weak fluctuations $\phi^i\ll 1/n_G$ lead to gauge
transformations and the constraints
\begin{equation}
\label{G1C1}
 \th_{ij}\to \th_{ij}+ \prt_i \th_j+\prt_j \th_i,\ \ \ \ \
 \prt_j \phi^{ji}=0
\end{equation}

The second  helicity $0$ mode is described by $\phi^{ij}=
(\del_{ij}\prt^2-\prt_i\prt_j) \phi$ and $\th=
(\del_{ij}\prt^2-\prt_i\prt_j)\th_{ij}$.  Its frequency is again $\om_{\v
k}=0$. The Hamiltonian for such a mode contains only $\th$.  So the quantum
fluctuations satisfies $ \del \phi \gg 1$ and $\del \th\ll 1/n_G$. The second
helicity $0$ mode is also gapped.  The strong fluctuations $\del \phi^{ij}=
(\del_{ij}\prt^2-\prt_i\prt_j) \phi \gg 1$ and the weak fluctuations $ \th=
(\del_{ij}\prt^2-\prt_i\prt_j)\th_{ij} \ll 1/n_G$ lead to a gauge
transformation and a constraint
\begin{equation}
\label{G2C2}
\phi^{ij}\to \phi^{ij}+(\del_{ij}\prt^2-\prt_i\prt_j) \phi,\ \ \ \ \
(\del_{ij}\prt^2-\prt_i\prt_j)\th_{ij}=0
\end{equation}

The Lagrangian \eq{Lagthphi} equipped with the gauge transformations and the
constraints (\ref{G1C1},\ref{G2C2}) is nothing but the linearized Einstein
Lagrangian of gravity, where $\th_{ij}\sim g_{ij}-\del_{ij}$ represents the
fluctuations of the metric tenor $g_{ij}$ around the flat space. So the
linearized Einstein gravity emerge from the quantum model \eq{HUJg} in the
large $n_G$ limit.  The local bosonic model \eq{HUJg} can be viewed as a
quantum theory of gravity.

We have seen that the gapping of the helicity $0$ mode in the rotor
model \eq{lattLC} leads to an emergence of $U(1)$ gauge structure at
low energies. The emergence of a gauge structure also represents a
new kind of order -- quantum order\cite{Wqoslpub,Wqoem} -- in the
ground state. In \Ref{HWcnt}, it was shown that
the emergent $U(1)$ gauge invariance, and hence the quantum order,
is robust against any local perturbations of the rotor model.  Thus
the gaplessness of the emergent photon is protected by the quantum
order.\cite{Wlight}  Similarly, the gapping of the two helicity $0$
modes and the helicity $\pm 1$ modes in the bosonic model \eq{HUJg}
leads to an emergent gauge invariance of the linearized coordinate
transformation. This indicates that the ground state of the bosonic
model contains a new kind of quantum order that is different from
those associated with emergent ordinary gauge invariances of
internal degrees of freedom.  We expect such an emergent linearized
diffeomorphism invariance to be robust against any local perturbation
of the bosonic model. Thus the gaplessness of the emergent gravitons is
protected by the quantum order.

The emergent gravitons in the model \eq{LagSp} naturally interact but the
interaction may be different from that described by the higher order
non-linear terms in Einstein gravity.  However, those higher order terms are
irrelevant at low energies.  Thus it may be possible to generate those higher
order terms by fine tuning the lattice model \eq{LagSp}, such as modifying the
Hamiltonian ($H_J$ and $H_g$), the constraints ($H_U$), as well as the Berry's
phase term in \eq{LagSp}.  So it may be possible that local bosonic models can
generate proper non-linear terms to satisfy (f).\cite{GWtoapp}

Our result appears to contradict with the Weinberg-Witten theorem\cite{WW8059}
which states that in all theories with a Lorentz-covariant energy-momentum
tensor, composite as well as elementary massless particles with helicity $h>1$
are forbidden.  However, the energy-momentum tensor in our model is not
invariant under the linearized diffeomorphism (although the action is
invariant). This may be the reason why emergent gravitons are possible in our
model.

We would like to thank J. Polchinski and E. Witten for their very helpful
comments.  This research is supported by NSF grant No.  DMR-0433632 and ARO
grant No. W911NF-05-1-0474.


\end{document}